\documentclass[letter]{aa}  
\usepackage{graphicx}
\usepackage{txfonts}
\usepackage{natbib}

\usepackage[FIGTOPCAP]{subfigure}
\usepackage{hyperref}
\hypersetup{
    colorlinks   = true,
    citecolor=blue,
}

\def\Hii{H\,{\sc ii}}

\def\kms{km\,s$^{-1}$}

\def\msun{M$_{\odot}$}
\def\msol{M$_{\odot}$}

\def\srv{$\sigma_\mathrm{RV}$}
\def\sdyn{$\sigma_\mathrm{dyn}$}

\def\Msol{M$_\odot$}
\def\Msun{M$_\odot$}

\def\xsh{X-shooter}
\def\srv{$\sigma_\mathrm{1D}$}
\def\fbin{$f_\mathrm{bin}$}

\def\pcutoff{$P_\mathrm{cutoff}$}

\def\ngc6357{NGC\,6357}
\def\g333{G333.6-0.2}

\def\minagem{1}
\def\maxagem{3}
\def\minagengc{0.5}
\def\maxagengc{3}

\def\maxageg{3}

\begin{document} 

   \title{A relation between the radial velocity dispersion of young clusters and their age:}
   \subtitle{Evidence for hardening as the formation scenario of massive close binaries}

  \author{\mbox{M.C. Ram\'irez-Tannus}\inst{1} 
         \and
         \mbox{F. Backs}\inst{2}
         \and
         \mbox{A. de Koter}\inst{2,3}
         \and
         \mbox{H. Sana}\inst{3}
         \and
         \mbox{H. Beuther}\inst{1}
         \and
         \mbox{A. Bik}\inst{4}
         \and
         \mbox{W. Brandner}\inst{1}
         \and
         \mbox{L. Kaper}\inst{2}
         \and
         \mbox{H. Linz}\inst{1}
         \and
         \mbox{Th. Henning}\inst{1}
         \and 
         \mbox{J. Poorta}\inst{2}
}

  \institute{
            Max Planck Institute for Astronomy, Königstuhl 17, 
            D-69117 Heidelberg, Germany;\\
             \email{ramirez@mpia.de}
        \and
        Astronomical Institute "Anton Pannekoek", University of Amsterdam,
             Science Park 904, 1098 XH Amsterdam, The Netherlands    
                \and
                    Institute of Astronomy, 
                    KU Leuven, Celestijnenlaan 200 D, 3001 Leuven, Belgium
        \and
                    Department of Astronomy, Stockholm University, 
                    Oskar Klein Center, SE-106 91 Stockholm, Sweden
            }
   \date{\today}

\abstract{
The majority of massive stars ($>8$\,\msol) in OB associations are found in close binary systems. Nonetheless, the formation mechanism of these close massive binaries is not understood yet. Using literature data, we measured the radial-velocity dispersion (\srv) as a proxy for the close binary fraction in ten OB associations in the Galaxy and the Large Magellanic Cloud, spanning an age range from 1 to 6~Myrs. We find a positive trend of this dispersion with the cluster's age, which is consistent with binary hardening. Assuming a universal binary fraction of \fbin = 0.7, we converted the \srv\ behavior to an evolution of the minimum orbital period \pcutoff\ from $\sim$9.5~years at 1~Myr to $\sim$1.4~days for the oldest clusters in our sample at $\sim$6~Myr.   
Our results suggest that binaries are formed at larger separations, and they harden in around 1 to 2~Myrs to produce the period distribution observed in few million year-old OB binaries. Such an inward migration may either be driven by an interaction with a remnant accretion disk or with other young stellar objects present in the system. 
Our findings constitute the first empirical evidence in favor of migration as a scenario for the formation of massive close binaries.
}


  \maketitle
%

\section{Introduction}

It is well established that the vast majority of massive stars ($M>8$~\msun) come in pairs or as higher-order multiples \citep[e.g.,][]{2009AJ....137.3358M,  2012MNRAS.424.1925C, 2012A&A...538A..74P, 2012ApJ...751....4K, 2014ApJS..213...34K, 2014ApJS..215...15S, 2015A&A...580A..93D}. A large fraction of these binaries have orbital periods on the order of 2 months or shorter \citep{2011IAUS..272..474S, 2012Sci...337..444S,2012ApJ...751....4K,2017A&A...598A..84A,barba_gamen_arias_morrell_2017}. These binaries are efficiently detected with spectroscopic techniques measuring periodic Doppler shifts of the photospheric lines. Massive binaries produce a variety of exotic products later in their evolution such as X-ray binaries, rare types of supernovae \citep[Ibc, IIn, super-luminous SNe,][]{1973ApJ...186.1007W, 2010ApJ...725..940Y, 2012ARA&A..50..107L}, gamma-ray bursts \citep[][]{1993ApJ...411..823W, 2007A&A...465L..29C}, and, eventually, gravitational wave sources \citep[e.g.,][]{2013A&ARv..21...59I, 2016MNRAS.458.2634M, 2016MNRAS.460.3545D, 2016MNRAS.462.3302E}. 
However, the origin of massive close binaries remains unknown.

The first effort to characterize the binarity properties of a sample of O stars in compact \Hii\ regions was performed by \citet{2007ApJ...655..484A}. They did a multi-epoch (two to three epochs) radial velocity (RV) study of a sample of 16 embedded O stars in seven massive star-forming regions. They identified two close binary stars based on their RV variations ($\sim90$~\kms) and measured an RV dispersion (\srv) of 35~\kms\ for the whole sample and 25~\kms\ when excluding the two close binaries.  

After pioneering studies to spectroscopically characterize single massive young stellar objects (mYSOs) such as those carried out by \citet{2006A&A...455..561B, 2012ApJ...744...87B}, \citet{2011A&A...536L...1O}, and \citet{2013A&A...558A.102E},  \citet{paper1} performed a single-epoch VLT/\xsh\ spectroscopic study of a sample of eleven candidate mYSOs in the very young giant \Hii\ region M17 ($\lesssim 1$ Myr). The stars range in mass from $6-25$~\Msol\ which is the mass range that dominates the samples from which multiplicity characteristics of 2-4~Myr old main-sequence OB stars are derived \citep{2012Sci...337..444S, 2014ApJS..213...34K}. The measured radial-velocity dispersion of these mYSOs is $\sigma_{\rm 1D} = 5.6\pm0.2$~\kms. In low density clusters, such as M17, \srv\ of a single epoch is strongly dominated by the orbital properties of the binary population. For example, if a given cluster has several close binaries of similar masses, one would expect the individual radial velocities of the stars to differ significantly from each other and, therefore, for \srv\ to be large. For 2-4~Myr clusters, with binary fractions $>0.5$ and minimum periods of $\sim1.4$~days, a dispersion of 30 to 50~\kms\ is typical \citep[e.g.,][]{2007A&amp;A...474...77K, 2008MNRAS.386..447S, 2012Sci...337..444S, 2014ApJS..211...10S, 2014ApJS..213...34K}. The latter is in stark contrast with our observation of M17, suggesting a lack of close massive binaries in this region.

In \citet{2017A&A...599L...9S}, two scenarios are explored that may explain the low \srv\ observed in M17: a small binary fraction \fbin\ and/or a lack of short-period binaries. They conclude that the observed dispersion can be explained either if $f_{\rm bin}=0.12^{+0.16}_{-0.09}$ or if the minimum orbital period $P_\mathrm{cutoff}>9$~months. Parent populations with $f_\mathrm{bin}>0.42$ or $P_\mathrm{cutoff}<47$~days can be rejected at the 95\% significance level. 
Since it is unlikely that the binary fraction for M17 would be so far below that of other clusters, this very interesting result suggests that massive binaries form in wide orbits that migrate inward over the course of a few million years. 
In this letter, we refer to the generic mechanism of shrinking binary periods as the migration scenario. 
One strong test for this scenario is to compare the velocity dispersion observed in clusters spanning a range of ages. If the binary orbits harden with time, one would expect \srv\ to increase as the cluster age increases. 

In \citet[][]{2020AA...633A.155R}, VLT/KMOS spectra of around 200 stars in three very young clusters (M8, \ngc6357, and \g333) were obtained. Introducing an automatic method to classify the spectra, the effective temperatures, and luminosities of the observed stars were characterized in order to place them in the Hertzsprung-Russell diagram (HRD). The age and mass range of the observed populations was constrained by comparison to MESA evolutionary tracks obtained from the MIST project \citep{2011ApJS..192....3P, 2013ApJS..208....4P, 2015ApJS..220...15P, 2016ApJS..222....8D, 2016ApJ...823..102C}. The main sequence stars in M8 have masses between $\sim 5$ and $\sim 70$~\msun\ and the age of this cluster is between \minagem\ and \maxagem~Myr. In \g333,\ the main sequence population ranges in mass between $\sim 5$ and $\sim 35$~\msun\ and the estimated age of this region is $< \maxageg$~Myr. The main sequence stars in \ngc6357\ have masses between $\sim 10$ and $\sim 100$~\msun\ and their ages range from $\minagengc-\maxagengc$~Myr. 

The goal of this paper is to provide a first test of the migration scenario for the formation of massive close binaries. We aim to study a possible age evolution of \srv, and \pcutoff, to constrain a timescale for binary hardening assuming a universal binary fraction \citep[\fbin$=0.7$;][]{2012Sci...337..444S}. We base our analysis on clusters younger than 6~Myr to ensure that neither secular evolution nor the effect of binary interactions \citep{1999A&A...350..148W, 2007A&A...467.1181D} affect our results significantly. 
In Section~\ref{P5:sec:RVdisp} we measure the radial velocities of the high-mass stars in M8 and \ngc6357\ and calculate their \srv. 
Next, we compare our findings with those presented by \citet{2012Sci...337..444S} for Galactic clusters of 2-4~Myr, with those from \citet{2018AJ....156..211Z} for Westerlund 2 (Wd2), with those from \citet{2012AA...546A..73H} for R136 in the Large Magellanic Cloud, and with those from \citet{paper1} for the very young massive-star forming region M17. This reveals a temporal behavior of \srv\ (Section~\ref{P5:sec:DispVsAge}) that is converted into an evolution of the minimum binary period (Section~\ref{sec:physical_parameters}), as binary motion is dominating the velocity dispersion of young massive clusters. 
In Section~\ref{P5:sec:Discussion} we discuss and conclude this work.


\section{Observations}
\label{P5:sec:RVdisp}

The sample studied in this paper consists of the OB stars in M8 and \ngc6357. The data acquisition and reduction are described in detail in \citet{2020AA...633A.155R}. In short, we obtained around 200 $H$ and $K$-band intermediate resolution spectra (with spectral resolution power, $\lambda/\Delta\lambda$, between 6700 and 8500, i.e., $30 < \Delta v < 40$~\kms) of stars in the abovementioned giant \Hii\ regions with VLT/KMOS \citep{2013Msngr.151...21S}. The final samples of massive stars consist of 16 stars in M8, 22 in \ngc6357, and four in \g333. We discarded \g333\ from our analysis because there are not enough stars with RV measurements to calculate \srv. 
A description of the age and mass range determination can be found in \citet{2020AA...633A.155R}, and Appendix~\ref{P5:sec:AccuracyAge} presents a detailed discussion about the accuracy of the age determination. 

The radial velocity (RV) of the intermediate to high-mass stars was obtained by measuring the Doppler shifts of a suitable set of  
photospheric lines. Tables~\ref{P5:tab:SpecLines_M8} and \ref{P5:tab:SpecLines_NGC6357} list the RV obtained for each star together with its error and the spectral lines used in our analysis.

The RV-fitting approach is similar to the one adopted 
by \citet[][]{2017A&A...599L...9S}. First, for the profile fitting, we adopted Gaussian profiles. Second, we clipped the core of diagnostic lines that were still contaminated by residuals of the nebular emission. Third, we simultaneously fit all spectral lines available, thereby assuming that the Doppler shift is the same for all lines \citep[see Section 2 and Appendix B of][]{2013A&A...550A.107S}.

Figure~\ref{P5:fig:RV_histograms} shows the radial-velocity distribution for the two regions. We calculated the errors of the histogram bins by randomly drawing RV values from a Gaussian centered at each measured RV and with a sigma corresponding to the measurement error; we repeated that process $10^{5}$ times. The value shown for each bin is the mean of all the RVs in that bin's measurements and the error bar corresponds to the standard deviation. 
We obtained \srv\ by calculating the weighted standard deviation of the measured RVs. 
The weighted mean and standard deviation are listed in the top-left corner of each histogram in Fig.~\ref{P5:fig:RV_histograms}. 
The measured \srv\ for M8 and \ngc6357\ are $32.7\pm2.6$ and $26.9\pm1.3$\,\kms,  respectively.


\section{Velocity dispersion versus cluster age}
\label{P5:sec:DispVsAge}

Based on single-epoch radial-velocity measurements of young massive stars in M17, \citet{2017A&A...599L...9S} conclude that this young ($\sim$1~Myr) star-forming region hosts only a few close binary systems. This is in contrast to the observation that most massive stars in somewhat older clusters are in close binaries. 
Under the hypothesis that massive stars form in wide binary systems that harden their orbits within the first million years of evolution, one would expect that during those initial couple of million years the radial-velocity dispersion, \srv,  increases with time. 

We plotted \srv\ of the clusters studied in this paper and compare it with the results by \citet[NGC\,6231, IC\,2944, IC\,1805, IC\,1848, NGC\,6611]{2012Sci...337..444S}, \citet[Wd2]{2018AJ....156..211Z}, \citet[R136]{2012AA...546A..73H}, and \citet[M17]{2017A&A...599L...9S}. R136 and Wd2 are very relevant for our study given their relatively young age (1-2~Myr) which is in between the age measured for M17 \citep{paper1} and that of the somewhat older clusters \citep{2012Sci...337..444S}. 
In order to compare the multi-epoch RV data provided by \citet{2012Sci...337..444S} with the single epoch data of M17, M8, \ngc6357, and Wd2, we drew the RV measured for each star in a given cluster in a random epoch and we computed the RV dispersion. We repeated this procedure $10^5$ times and then calculated the most probable \srv\ and its standard deviation. The \srv\ obtained for each cluster is listed in the third column of Table~\ref{P5:tab:RVdisp_vs_age}. The second and seventh columns show the age of the clusters and the respective references.

\begin{table}
\caption{Age, radial-velocity dispersion, number of stars, and mass range for our sample of young clusters hosting massive stars.} 
\label{P5:tab:RVdisp_vs_age} 
\centering
\renewcommand{\arraystretch}{1.4}
\setlength{\tabcolsep}{2pt}
\begin{tabular}{lcccccc}
\hline
Cluster & Age & $\sigma_{1D}$ & N & Mass & $P_{\rm min}$ & Age \\
 & Myr & \kms\ &  stars & \msun & days &  ref. \\
\hline
IC1805 & 1.6 -- 3.5 & $65.5\pm3.1$ & 8 & 15 -- 60 & $1.4^{+0.8}$ & 1 \\ 
IC1848 & 3.0 -- 5.0 & $50.3\pm12.8$ & 5 & 15 -- 60 & $1.4^{+3.2}$ & 2 \\ 
IC2944 & 2.0 -- 3.0 & $31.4\pm0.3$ & 14 & 15 -- 60 & $6.3_{-4.5}^{+17.8}$ & 3 \\
NGC6231 & 3.5 -- 5.4 & $67.6\pm0.4$ & 13 & 15 -- 60 & $1.4^{+0.7}$ & 4 \\
NGC6611 & 2.0 -- 6.0 & $25.3\pm1.6$ & 9 & 15 -- 60 & $10.1_{-8.5}^{+43.0}$ & 5 \\
Wd2 & 1.0 -- 2.0 & $15.0\pm0.1$ & 44 & 6 -- 60 & $62.2_{-36.1}^{+74.9}$ & 6 \\
M17 & 0.0 -- 2.0 & $5.5\pm0.5$ & 12 & 6 -- 20 & $3500_{-2834}$ & 7 \\
M8 & 1.0 -- 3.0 & $32.7\pm2.6$ & 16 & 6 -- 20 & $2.4_{-1.0}^{+3.8}$ & 8 \\
NGC6357 & 0.5 -- 3.0 & $26.9\pm1.3$ & 22 & 6 -- 30 & $5.4_{-3.4}^{+8.5}$ & 8 \\
R136 & 1.0 -- 2.0 & $25.0\pm5.9$ & 332 & 15 -- 60 & $24.1_{-5.1}^{+9.0}$ & 9 \\
\hline
\end{tabular}
\tablebib{(1)~\citet{2017ApJS..230....3S}; (2)~\citet{2014MNRAS.438.1451L}; (3)~\citet{2014MNRAS.443..411B}; (4)~van der Meij et al. (2020), \textit{in prep.}; (5)~\citet{2008AA...490.1071G}; (6)~\citet{2018AJ....156..211Z}; (7)~\citet{paper1}; (8)~\citet{2020AA...633A.155R}; (9)~\citet{2012AA...546A..73H}}
\end{table}

\begin{figure}
\includegraphics[width=0.93\hsize]{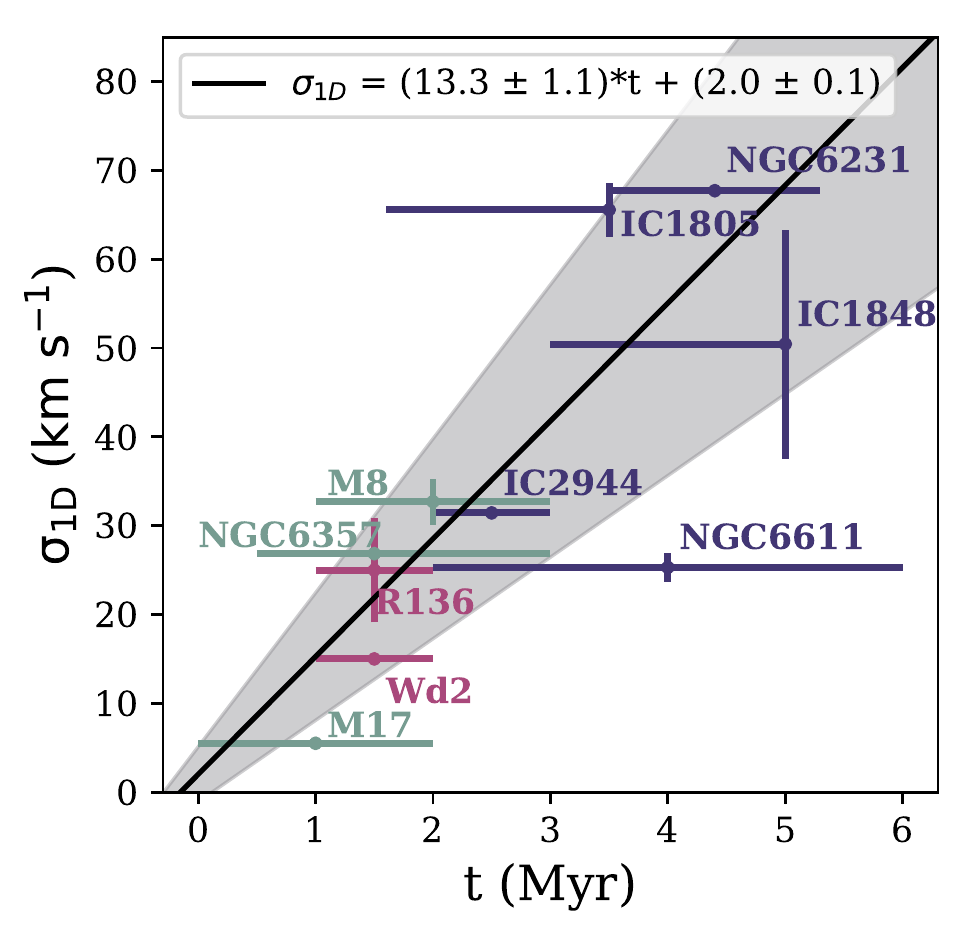}
  \caption{Radial-velocity dispersion (\srv) versus age of the clusters. The purple data points show the data from \citet{2012Sci...337..444S}, the magenta points show Wd2 and R136 \citep{2018AJ....156..211Z, 2012AA...546A..73H}, and the green data points show the clusters studied in \citet{2020AA...633A.155R} and \citet{2017A&A...599L...9S}. The solid black line represents the linear fit to the data and the gray area shows the 1-$\sigma$ errors on the fit. 
}
     \label{P5:fig:RVdisp_vs_age}
\end{figure}

\begin{figure}
    {\includegraphics[width=\hsize]{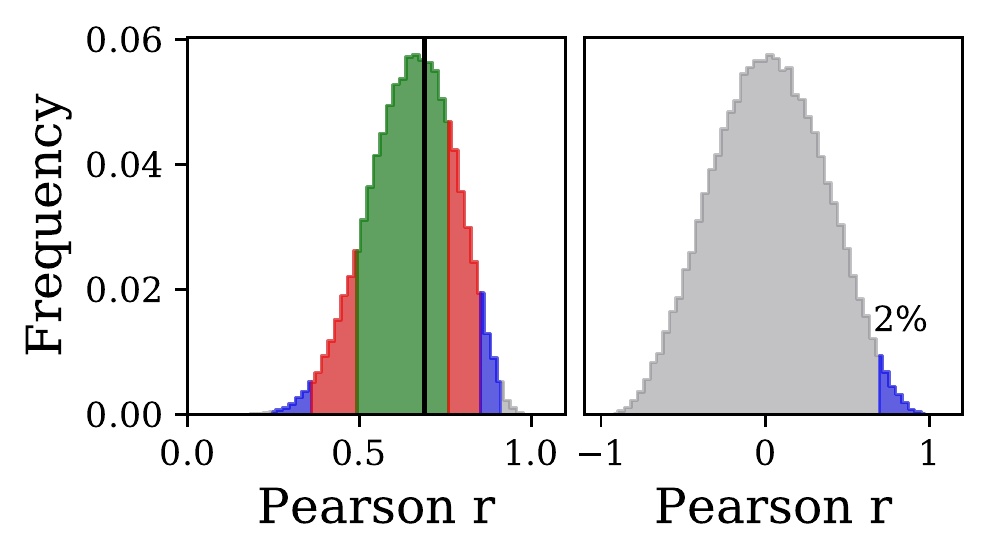}}
  \caption{Distribution of Pearson coefficients after randomly drawing $10^5$ samples from the data shown in Figure~\ref{P5:fig:RVdisp_vs_age}. \emph{Left:} 2D Gaussians centered at our data points and with $\sigma$ equal to our error bars. The green, red, and blue areas show the 68, 95, and 99\% confidence intervals, respectively. The black line shows the observed coefficient. \emph{Right:} 2D Gaussians centered at random locations of the parameter space (age between 0 and 7~Myr and \srv\ between 0 and 80~\kms) and with $\sigma$ equal to our error bars. The blue shaded area represents the probability (2\%) that the observed coefficient is caused by a random distribution.}
     \label{P5:fig:PearsonTests}
\end{figure}

In Figure~\ref{P5:fig:RVdisp_vs_age} we show \srv\ versus age of the clusters. We performed an orthogonal distance regression \citep[ODR;][]{1990odr_reference} to the data and find a positive correlation between the age of the clusters and \srv. 
The solid black line represents the best fit to the data and the gray area represents the 1-$\sigma$ error on the fit. 
The Pearson coefficient for the observed relation is 0.7, which indicates a strong positive correlation. Nevertheless, this coefficient does not take into account the errors in the parameters. To test the validity of our results we performed two Monte Carlo tests whose results are shown in Figure~\ref{P5:fig:PearsonTests}. The left panel shows the distribution of Pearson coefficients obtained from drawing random points centered on our data (age, \srv) with a standard deviation equal to our error bars. The right panel shows the probability (2\%) that a random distribution causes the observed coefficient.

Even though the scatter is substantial, we conclude that for the present data there is a positive correlation between \srv\ and the age of the clusters. Although the number of clusters considered here remains limited and there are sizable uncertainties as to the individual age determinations, our results indicate an increase in the fraction of close binary systems as the clusters get older. M17 seems to be a unique cluster given its very young age and extremely low \srv. Trumpler\,14 in this sense seems comparable, being $\sim$1~Myr old and having no evidence for close binaries amongst its O stars \citep{2011IAUS..272..474S}. We do not include this cluster in the analysis because there are not enough stars with RV measurements to calculate \srv. 
 
\section{Physical parameters}
\label{sec:physical_parameters}

Our aim is to characterize the multiplicity properties of observed clusters based on the observed \srv. This section presents the results of Monte Carlo population syntheses computed with different underlying multiplicity properties.

\subsection{Binary fraction and minimum period}

 We focus on the effect that the binary fraction, \fbin, and minimum orbital period, \pcutoff, have on the observed \srv. The methodology is similar to that used in \citet{2017A&A...599L...9S}. A parent population of stars is generated with a certain \fbin\ and \pcutoff. The binary star systems in this population are described by their orbital properties. These are the primary mass, $M_1$, mass ratio, $q$, period, $P$, and eccentricity, $e$. For the multiplicity properties that we do not vary, we adopt the values from \citet{2012Sci...337..444S} for Galactic young clusters as the basic properties for our population. For the primary star, we adopt a Kroupa mass distribution \citep{2001MNRAS.322..231K}. The mass ratio distribution is uniform with $0.1 < q < 1$. The probability density function of the period is described as $\text{pdf}(P) \propto (\log P)^{-0.5}$, with the period in days and $\log P_{\rm cutoff} < \log P < 3.5$. The eccentricity distribution depends on the period of the binary system. For $P<4$ days, we assume circular orbits; for periods between 4 and 6 days, the eccentricities are sampled from $\text{pdf}(e) \propto e^{-0.5}$, with $0 \leq e < 0.5$; for periods longer than 6 days the same distribution is used, but with $0 \leq e < 0.9$. 

\begin{figure}
\includegraphics[width=\hsize]{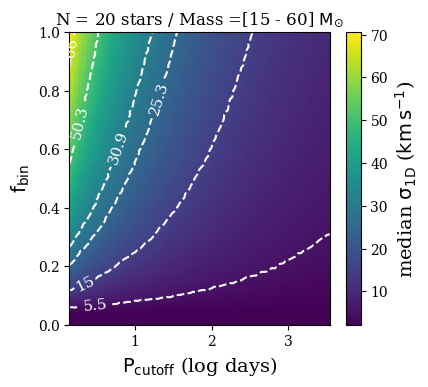}
\caption{Median \srv\ obtained for distributions with different combinations of \fbin\ and \pcutoff. For this set of simulations, we used a sample of 20 stars ranging in masses from 15 to 60~\msun. The contours show the trends followed by \srv\ values of 5.5, 15, 25.3, 30.9, 50.3, and 66 \kms. 
}
\label{P5:fig:MonteCarlo}
\end{figure}

In order to calculate the radial velocity of the binary star systems, they are given a randomly generated inclination, $i$, longitude of periastron, $\omega$, and eccentric anomaly, $E$. The latter is determined by generating a random mean anomaly and numerically solving Kepler's equation to find the corresponding eccentric anomaly using Brent's method \citep{1973SJNA...10..327B}\footnote{\url{https://docs.scipy.org/doc/scipy/reference/generated/scipy.optimize.brentq.html}}. 
This gives all of the information required to calculate the binary component of the radial velocity of the primary star through
\begin{equation}
    v_{\rm r,b} = K_1\left(e \cos(\omega) + \cos\left[2 \arctan\left(\sqrt{\frac{1 + e}{1 - e}} \tan\left(\frac{1}{2}E\right)\right) + \omega\right] \right),
\end{equation}
with $K_1$ being the amplitude of the orbital velocity of the primary star. The binary component of the radial velocity is added to the velocity due to cluster dynamics, \sdyn. This results in a population of stars with radial velocities based on either cluster dynamics only (in the case of \fbin=0) or both cluster dynamics and binary orbits. The secondary stars are assumed to be undetected. We adopt a cluster velocity dispersion for all clusters of $2\,{\rm km\,s^{-1}}$, which corresponds to the typical value found by  \cite{2019ApJ...870...32K} who measured \sdyn\ for several massive clusters in the Milky Way. Adopting \sdyn~$=1$ or \sdyn$=5$~\kms\ has no significant effect on our results. 
\citet{2012Sci...337..444S} also conclude that the dynamical dispersion of young clusters is negligible with respect to the dispersion due to binary motions. 
Each Monte Carlo run consists of a generated parent population of $10^5$ stars, which is sampled $10^5$ times to simulate the observation of a cluster. The mass distribution of the stars in the parent distribution are matched to the mass distribution in the observed cluster. The samples contain the same number of stars and measurement accuracy as the observed sample. We then constructed density distributions of \srv\ for each parent population. 
Each grid point in Figure~\ref{P5:fig:MonteCarlo} shows the median \srv\ obtained for a simulated cluster of 20 stars sampled from a parent population with masses between 15 and 60~\msun, where we varied \fbin, and \pcutoff. The color bar corresponds to \srv\ in \kms. The dashed lines show the \srv\ contours for 5.5, 15, 25.3, 30.9, 50.3, and 66 \kms, which correspond to the \srv\ measured for our clusters. 
As the number of stars and the mass ranges are different for each cluster, the contours do  not  represent  the  exact  way  in  which  we  measured \pcutoff\ but are meant to show examples of the trends that a certain \srv\ follows in this diagram. A comparison of our results with \citet{2017A&A...599L...9S} is shown in Appendix~\ref{P5:sec:PrevWork}.

\subsection{Timescale of binary hardening}

\begin{figure}
\includegraphics[width=0.93\hsize]{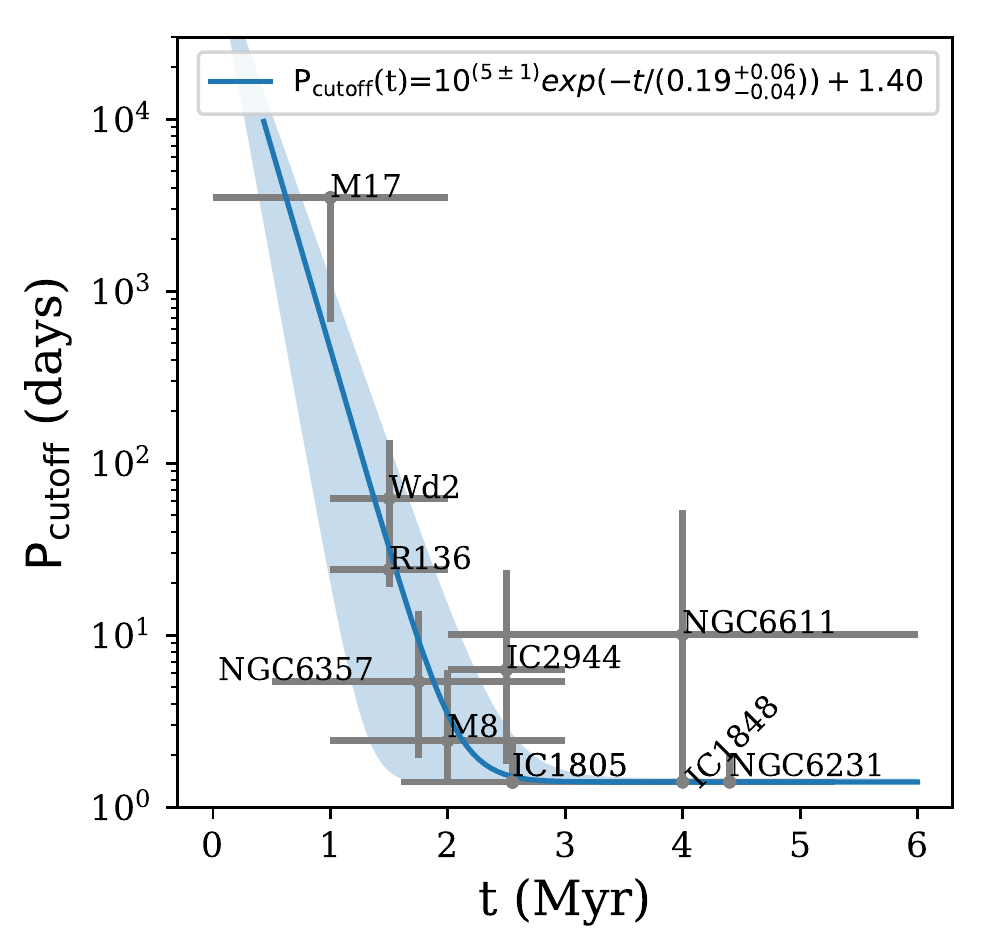}
  \caption{$P_{\rm cutoff}$ versus age of the clusters. The error bars represent the \pcutoff\ corresponding to the distributions that represent each \srv\ within their 68\% confidence range. The blue line and shaded region show the fit to the data and its 1-$\sigma$ error bars.}
     \label{P5:fig:Pcutoff_vs_age}
\end{figure}

Assuming that the binary fraction of massive stars is consistent with that observed in OB stars in young open clusters \citep[$f_{\rm bin}=0.7$;][]{2012Sci...337..444S}, but that binary stars are born in wider orbits, we can estimate \pcutoff\ which best represents the observed \srv\ for each cluster. For each cluster, we kept the distribution of orbital properties fixed and we adjusted the sample size and mass ranges in accordance to those of the observed samples (see Table~\ref{P5:tab:RVdisp_vs_age}). This allowed us to make a similar plot as in Figure~\ref{P5:fig:RVdisp_vs_age} but in terms of \pcutoff. Figure~\ref{P5:fig:Pcutoff_vs_age} shows the estimated \pcutoff\ for each cluster as a function of cluster age. 
We determined that, assuming $f_\mathrm{bin}=0.7$, the most likely \pcutoff\ to explain the observed \srv\ of M17 is around 3500~days and that a \pcutoff\ of  665~days and 126~days can be rejected at the 68\% and 95\% confidence levels, respectively. For the clusters with the largest \srv\ in our sample, IC\,1805, IC\,1848, and NGC\,6231, the cutoff period that best explains the observed \srv\ is 1.4~days, which is the \pcutoff\ adopted by \citet{2012Sci...337..444S}. 

In order to get a first estimate of the binary hardening timescale, we fit a function of the form $P(t) = P_{0}e^{-t/t_{0}} + c$ to the data in Figure~\ref{P5:fig:Pcutoff_vs_age}, where $P_0$ is the minimum period at the moment of binary formation and $t_0$ corresponds to the e-folding time. As $P_0$ is very uncertain, we assumed a typical value of $10^5$~days, corresponding to an initial separation of $\sim$100~AU for a pair of 10~\msun\ stars. In order to find the range of $t_0$, we varied $P_0$ from $10^4$ to $10^6$~days . The resulting fit and corresponding parameter ranges are shown with the blue line and shaded area in Figure~\ref{P5:fig:Pcutoff_vs_age}. We obtain a typical e-folding time of $t_0 = 0.19_{-0.04}^{+0.06}$~Myr. To harden an orbit from 3500 to 1.4~days, $\sim$8 e-foldings are needed, this implies that a typical binary hardening timescale is $\sim$1.6~Myr. Assuming a binary system with two 10~\msun\ stars, this would imply a mean hardening rate of $\sim7.5~{\rm AU~ yr^{-1}}$. For M17, we discarded periods lower than 665~d with a 68\% confidence level. Following the same argument as before, the typical timescale to harden an orbit from 665 to 1.4~days would be $\sim$1~Myr equivalent to a mean hardening rate of $\sim4~{\rm AU~ yr^{-1}}$. 

\section{Discussion and conclusions}
\label{P5:sec:Discussion}

In this paper we present observational evidence that the 1D velocity dispersion of massive stars in young clusters (\srv) increases as they get older (Figure~\ref{P5:fig:RVdisp_vs_age}). Additionally, we performed Monte Carlo simulations which allowed us to convert the measured values of \srv\ to physical parameters. Assuming that stars are born with the binary fraction representative of OB stars in $2-4$~Myr clusters (\fbin=0.7), we calculated the cutoff period that would correspond to each of the observed populations. From Figure~\ref{P5:fig:Pcutoff_vs_age} we can conclude that the orbits would harden in 1-1.5~Myrs. It is worth pointing out that the fit relies heavily on the M17 point. If the binary fraction of this cluster were lower than 0.7, the timescales derived would be larger. Nevertheless, the qualitative results of this paper remain the same.

\citet{2017A&A...599L...9S} propose that the inward migration process could be driven either by the interaction with the remnants of an accretion disk or with other young stellar objects. The hardening of the orbits would then stop either when the disk is destroyed, or when the third body is pushed far out or ejected from the orbit. For a typical system of two 10~\Msun\ stars, a total angular momentum of $\sim6.5\times10^{47}~{\rm kg~m^{2}~s^{-1}}$ ($\sim3.5\times10^{47}~{\rm kg~m^{2}~s^{-1}}$) needs to be transferred for the binary to harden from 3500 (665) to 1.4 days. 
\citet{2019MNRAS.488.2480R} explored the possibility of the orbit hardening via the Eccentric Kozai-Lidov (EKL) mechanism, where a third companion perturbs the orbit of a binary system. They find that, beginning with a cutoff period of 9 months the EKL mechanism is insufficient to reproduce the population of short period binaries observed by \citet{2012Sci...337..444S}. They suggest that type II migration \citep{1986ApJ...309..846L} might explain the tightening of binaries in such a time period. 
\citet{2018ApJ...854...44M} find that the main mechanism to harden binaries should be the dynamical disruption of coplanar triples that initially fragmented in the disk in combination with energy dissipation within this disk. Other mechanisms include the combination of EKL oscillations with tidal friction both during the pre-main sequence and the main sequence \citep{2002MNRAS.336..705B, 2009MNRAS.392..590B}. Given the densities of the clusters studied, stellar encounters are uncommon and, therefore, mechanisms such as binary-binary or single-binary interactions should not play a significant role in the formation of close binaries \citep{2011Sci...334.1380F}.

Our findings affect predictions of binary population synthesis models that follow the evolution of an ensemble of binary systems subject to, among others, a distribution function for the zero-age orbital periods. Such models 
\citep[e.g.,][]{2012Sci...337..444S,2015ApJ...805...20S} predict that a small fraction of systems interact within a timescale of 3\,Myr, almost invariably resulting in a merging of the two components thus creating a blue straggler. Specifically, \citet{2015ApJ...805...20S} report that of the population within two magnitudes in brightness of the main-sequence turnoff, only 1\%, 2\%, and 10\% is a blue straggler after 1, 2, or 3 Myr. Therefore, unless one is specifically interested in early blue straggler formation, the temporal evolution of orbital properties in the first few million years reported here should not affect predictions made by binary population synthesis models that rely on initial conditions defined at the zero-age main sequence only. Except for the caveat mentioned, adopting initial conditions for the massive star population as reported on by \citet{2012Sci...337..444S}, for example, remains a justified approach.


\begin{acknowledgements}
Based on observations collected at the European Organisation for Astronomical Research in the Southern Hemisphere under ESO program 095.C-0048. 
This research made use of Astropy, a community-developed core Python package for Astronomy \citep{2018arXiv180102634T}, 
NASA's Astrophysics Data System Bibliographic Services (ADS), the SIMBAD database, operated at CDS, Strasbourg, France \citep{2000A&AS..143....9W}, and the cross-match service provided by CDS, Strasbourg. 
The research leading to these results has received funding from the European Research Council (ERC) under the European Union's Horizon 2020 research and innovation programme (grant agreement numbers 772225: MULTIPLES).

\end{acknowledgements}


\bibliographystyle{aa}
\bibliography{references}

\begin{thebibliography}{71}
\expandafter\ifx\csname natexlab\endcsname\relax\def\natexlab#1{#1}\fi

\bibitem[{{Almeida} {et~al.}(2017){Almeida}, {Sana}, {Taylor}, {Barb{\'a}},
  {Bonanos}, {Crowther}, {Damineli}, {de Koter}, {de Mink}, {Evans}, {Gieles},
  {Grin}, {H{\'e}nault-Brunet}, {Langer}, {Lennon}, {Lockwood}, {Ma{\'{\i}}z
  Apell{\'a}niz}, {Moffat}, {Neijssel}, {Norman}, {Ram{\'{\i}}rez-Agudelo},
  {Richardson}, {Schootemeijer}, {Shenar}, {Soszy{\'n}ski}, {Tramper}, \&
  {Vink}}]{2017A&A...598A..84A}
{Almeida}, L.~A., {Sana}, H., {Taylor}, W., {et~al.} 2017, \aap, 598, A84

\bibitem[{{Alonso} {et~al.}(1999){Alonso}, {Arribas}, \&
  {Mart{\'{\i}}nez-Roger}}]{1999A&AS..140..261A}
{Alonso}, A., {Arribas}, S., \& {Mart{\'{\i}}nez-Roger}, C. 1999, \aaps, 140,
  261

\bibitem[{{Apai} {et~al.}(2007){Apai}, {Bik}, {Kaper}, {Henning}, \&
  {Zinnecker}}]{2007ApJ...655..484A}
{Apai}, D., {Bik}, A., {Kaper}, L., {Henning}, T., \& {Zinnecker}, H. 2007,
  \apj, 655, 484

\bibitem[{{Arias} {et~al.}(2007){Arias}, {Barb{\'a}}, \&
  {Morrell}}]{2007MNRAS.374.1253A}
{Arias}, J.~I., {Barb{\'a}}, R.~H., \& {Morrell}, N.~I. 2007, \mnras, 374, 1253

\bibitem[{Barb\'a {et~al.}(2017)Barb\'a, Gamen, Arias, \&
  Morrell}]{barba_gamen_arias_morrell_2017}
Barb\'a, R.~H., Gamen, R., Arias, J.~I., \& Morrell, N.~I. 2017, Proceedings of
  the International Astronomical Union, 12, 89

\bibitem[{{Bate}(2009)}]{2009MNRAS.392..590B}
{Bate}, M.~R. 2009, \mnras, 392, 590

\bibitem[{{Bate} {et~al.}(2002){Bate}, {Bonnell}, \&
  {Bromm}}]{2002MNRAS.336..705B}
{Bate}, M.~R., {Bonnell}, I.~A., \& {Bromm}, V. 2002, \mnras, 336, 705

\bibitem[{{Baume} {et~al.}(2014){Baume}, {Rodr{\'{\i}}guez}, {Corti},
  {Carraro}, \& {Panei}}]{2014MNRAS.443..411B}
{Baume}, G., {Rodr{\'{\i}}guez}, M.~J., {Corti}, M.~A., {Carraro}, G., \&
  {Panei}, J.~A. 2014, \mnras, 443, 411

\bibitem[{{Bik} {et~al.}(2012){Bik}, {Henning}, {Stolte}, {Brandner},
  {Gouliermis}, {Gennaro}, {Pasquali}, {Rochau}, {Beuther}, {Ageorges},
  {Seifert}, {Wang}, \& {Kudryavtseva}}]{2012ApJ...744...87B}
{Bik}, A., {Henning}, T., {Stolte}, A., {et~al.} 2012, \apj, 744, 87

\bibitem[{{Bik} {et~al.}(2006){Bik}, {Kaper}, \&
  {Waters}}]{2006A&A...455..561B}
{Bik}, A., {Kaper}, L., \& {Waters}, L.~B.~F.~M. 2006, \aap, 455, 561

\bibitem[{{Brent}(1973)}]{1973SJNA...10..327B}
{Brent}, R.~P. 1973, SIAM Journal on Numerical Analysis, 10, 327

\bibitem[{{Cantiello} {et~al.}(2007){Cantiello}, {Yoon}, {Langer}, \&
  {Livio}}]{2007A&A...465L..29C}
{Cantiello}, M., {Yoon}, S.~C., {Langer}, N., \& {Livio}, M. 2007, \aap, 465,
  L29

\bibitem[{{Chini} {et~al.}(2012){Chini}, {Hoffmeister}, {Nasseri}, {Stahl}, \&
  {Zinnecker}}]{2012MNRAS.424.1925C}
{Chini}, R., {Hoffmeister}, V.~H., {Nasseri}, A., {Stahl}, O., \& {Zinnecker},
  H. 2012, \mnras, 424, 1925

\bibitem[{{Choi} {et~al.}(2016){Choi}, {Dotter}, {Conroy}, {Cantiello},
  {Paxton}, \& {Johnson}}]{2016ApJ...823..102C}
{Choi}, J., {Dotter}, A., {Conroy}, C., {et~al.} 2016, \apj, 823, 102

\bibitem[{{Churchwell}(1990)}]{1990odr_reference}
{Churchwell}, E. 1990, in Contemporary mathematics (American Mathematical
  Society), Vol. 112, Hot Star Workshop III: The Earliest Phases of Massive
  Star Birth, ed. P.~{Brown} \& W.~{Fuller}, 186

\bibitem[{{Cox}(2000)}]{2000asqu.book.....C}
{Cox}, A.~N. 2000, {Allen's astrophysical quantities}

\bibitem[{{de Mink} \& {Mandel}(2016)}]{2016MNRAS.460.3545D}
{de Mink}, S.~E. \& {Mandel}, I. 2016, \mnras, 460, 3545

\bibitem[{{de Mink} {et~al.}(2007){de Mink}, {Pols}, \&
  {Hilditch}}]{2007A&A...467.1181D}
{de Mink}, S.~E., {Pols}, O.~R., \& {Hilditch}, R.~W. 2007, \aap, 467, 1181

\bibitem[{Dormand \& Prince(1980)}]{dormand1980family}
Dormand, J.~R. \& Prince, P.~J. 1980, Journal of computational and applied
  mathematics, 6, 19

\bibitem[{{Dotter}(2016)}]{2016ApJS..222....8D}
{Dotter}, A. 2016, \apjs, 222, 8

\bibitem[{{Dunstall} {et~al.}(2015){Dunstall}, {Dufton}, {Sana}, {Evans},
  {Howarth}, {Sim{\'o}n-D{\'{\i}}az}, {de Mink}, {Langer}, {Ma{\'{\i}}z
  Apell{\'a}niz}, \& {Taylor}}]{2015A&A...580A..93D}
{Dunstall}, P.~R., {Dufton}, P.~L., {Sana}, H., {et~al.} 2015, \aap, 580, A93

\bibitem[{{Eldridge} \& {Stanway}(2016)}]{2016MNRAS.462.3302E}
{Eldridge}, J.~J. \& {Stanway}, E.~R. 2016, \mnras, 462, 3302

\bibitem[{{Ellerbroek} {et~al.}(2013){Ellerbroek}, {Bik}, {Kaper}, {Maaskant},
  {Paalvast}, {Tramper}, {Sana}, {Waters}, \& {Balog}}]{2013A&A...558A.102E}
{Ellerbroek}, L.~E., {Bik}, A., {Kaper}, L., {et~al.} 2013, \aap, 558, A102

\bibitem[{{Feigelson} {et~al.}(2013){Feigelson}, {Townsley}, {Broos}, {Busk},
  {Getman}, {King}, {Kuhn}, {Naylor}, {Povich}, {Baddeley}, {Bate},
  {Indebetouw}, {Luhman}, {McCaughrean}, {Pittard}, {Pudritz}, {Sills}, {Song},
  \& {Wadsley}}]{2013ApJS..209...26F}
{Feigelson}, E.~D., {Townsley}, L.~K., {Broos}, P.~S., {et~al.} 2013, \apjs,
  209, 26

\bibitem[{{Fujii} \& {Portegies Zwart}(2011)}]{2011Sci...334.1380F}
{Fujii}, M.~S. \& {Portegies Zwart}, S. 2011, Science, 334, 1380

\bibitem[{{Gvaramadze} \& {Bomans}(2008)}]{2008AA...490.1071G}
{Gvaramadze}, V.~V. \& {Bomans}, D.~J. 2008, \aap, 490, 1071

\bibitem[{{H{\'e}nault-Brunet} {et~al.}(2012){H{\'e}nault-Brunet}, {Evans},
  {Sana}, {Gieles}, {Bastian}, {Ma{\'{\i}}z Apell{\'a}niz}, {Markova},
  {Taylor}, {Bressert}, {Crowther}, \& {van Loon}}]{2012AA...546A..73H}
{H{\'e}nault-Brunet}, V., {Evans}, C.~J., {Sana}, H., {et~al.} 2012, \aap, 546,
  A73

\bibitem[{{Indebetouw} {et~al.}(2005){Indebetouw}, {Mathis}, {Babler}, {Meade},
  {Watson}, {Whitney}, {Wolff}, {Wolfire}, {Cohen}, {Bania}, {Benjamin},
  {Clemens}, {Dickey}, {Jackson}, {Kobulnicky}, {Marston}, {Mercer},
  {Stauffer}, {Stolovy}, \& {Churchwell}}]{2005ApJ...619..931I}
{Indebetouw}, R., {Mathis}, J.~S., {Babler}, B.~L., {et~al.} 2005, \apj, 619,
  931

\bibitem[{{Ivanova} {et~al.}(2013){Ivanova}, {Justham}, {Chen}, {De Marco},
  {Fryer}, {Gaburov}, {Ge}, {Glebbeek}, {Han}, {Li}, {Lu}, {Marsh},
  {Podsiadlowski}, {Potter}, {Soker}, {Taam}, {Tauris}, {van den Heuvel}, \&
  {Webbink}}]{2013A&ARv..21...59I}
{Ivanova}, N., {Justham}, S., {Chen}, X., {et~al.} 2013, \aapr, 21, 59

\bibitem[{{Kiminki} \& {Kobulnicky}(2012)}]{2012ApJ...751....4K}
{Kiminki}, D.~C. \& {Kobulnicky}, H.~A. 2012, \apj, 751, 4

\bibitem[{{Kobulnicky} {et~al.}(2014){Kobulnicky}, {Kiminki}, {Lundquist},
  {Burke}, {Chapman}, {Keller}, {Lester}, {Rolen}, {Topel}, {Bhattacharjee},
  {Smullen}, {Vargas {\'A}lvarez}, {Runnoe}, {Dale}, \&
  {Brotherton}}]{2014ApJS..213...34K}
{Kobulnicky}, H.~A., {Kiminki}, D.~C., {Lundquist}, M.~J., {et~al.} 2014,
  \apjs, 213, 34

\bibitem[{{Kouwenhoven} {et~al.}(2007){Kouwenhoven}, {Brown}, {Portegies
  Zwart}, \& {Kaper}}]{2007A&amp;A...474...77K}
{Kouwenhoven}, M.~B.~N., {Brown}, A.~G.~A., {Portegies Zwart}, S.~F., \&
  {Kaper}, L. 2007, \aap, 474, 77

\bibitem[{{Kroupa}(2001)}]{2001MNRAS.322..231K}
{Kroupa}, P. 2001, \mnras, 322, 231

\bibitem[{{Kuhn} {et~al.}(2019){Kuhn}, {Hillenbrand}, {Sills}, {Feigelson}, \&
  {Getman}}]{2019ApJ...870...32K}
{Kuhn}, M.~A., {Hillenbrand}, L.~A., {Sills}, A., {Feigelson}, E.~D., \&
  {Getman}, K.~V. 2019, \apj, 870, 32

\bibitem[{{Langer}(2012)}]{2012ARA&A..50..107L}
{Langer}, N. 2012, \araa, 50, 107

\bibitem[{{Lim} {et~al.}(2014){Lim}, {Sung}, {Kim}, {Bessell}, \&
  {Karimov}}]{2014MNRAS.438.1451L}
{Lim}, B., {Sung}, H., {Kim}, J.~S., {Bessell}, M.~S., \& {Karimov}, R. 2014,
  \mnras, 438, 1451

\bibitem[{{Lin} \& {Papaloizou}(1986)}]{1986ApJ...309..846L}
{Lin}, D.~N.~C. \& {Papaloizou}, J. 1986, \apj, 309, 846

\bibitem[{{Mandel} \& {de Mink}(2016)}]{2016MNRAS.458.2634M}
{Mandel}, I. \& {de Mink}, S.~E. 2016, \mnras, 458, 2634

\bibitem[{{Martins} \& {Palacios}(2013)}]{2013A&A...560A..16M}
{Martins}, F. \& {Palacios}, A. 2013, \aap, 560, A16

\bibitem[{{Mason} {et~al.}(2009){Mason}, {Hartkopf}, {Gies}, {Henry}, \&
  {Helsel}}]{2009AJ....137.3358M}
{Mason}, B.~D., {Hartkopf}, W.~I., {Gies}, D.~R., {Henry}, T.~J., \& {Helsel},
  J.~W. 2009, \aj, 137, 3358

\bibitem[{{Massi} {et~al.}(2015){Massi}, {Giannetti}, {Di Carlo}, {Brand},
  {Beltr{\'a}n}, \& {Marconi}}]{2015A&A...573A..95M}
{Massi}, F., {Giannetti}, A., {Di Carlo}, E., {et~al.} 2015, \aap, 573, A95

\bibitem[{{Moe} \& {Kratter}(2018)}]{2018ApJ...854...44M}
{Moe}, M. \& {Kratter}, K.~M. 2018, \apj, 854, 44

\bibitem[{{Nishiyama} {et~al.}(2009){Nishiyama}, {Tamura}, {Hatano}, {Kato},
  {Tanab{\'e}}, {Sugitani}, \& {Nagata}}]{2009ApJ...696.1407N}
{Nishiyama}, S., {Tamura}, M., {Hatano}, H., {et~al.} 2009, \apj, 696, 1407

\bibitem[{{Ochsendorf} {et~al.}(2011){Ochsendorf}, {Ellerbroek}, {Chini},
  {Hartoog}, {Hoffmeister}, {Waters}, \& {Kaper}}]{2011A&A...536L...1O}
{Ochsendorf}, B.~B., {Ellerbroek}, L.~E., {Chini}, R., {et~al.} 2011, \aap,
  536, L1

\bibitem[{{Paxton} {et~al.}(2011){Paxton}, {Bildsten}, {Dotter}, {Herwig},
  {Lesaffre}, \& {Timmes}}]{2011ApJS..192....3P}
{Paxton}, B., {Bildsten}, L., {Dotter}, A., {et~al.} 2011, \apjs, 192, 3

\bibitem[{{Paxton} {et~al.}(2013){Paxton}, {Cantiello}, {Arras}, {Bildsten},
  {Brown}, {Dotter}, {Mankovich}, {Montgomery}, {Stello}, {Timmes}, \&
  {Townsend}}]{2013ApJS..208....4P}
{Paxton}, B., {Cantiello}, M., {Arras}, P., {et~al.} 2013, \apjs, 208, 4

\bibitem[{{Paxton} {et~al.}(2015){Paxton}, {Marchant}, {Schwab}, {Bauer},
  {Bildsten}, {Cantiello}, {Dessart}, {Farmer}, {Hu}, {Langer}, {Townsend},
  {Townsley}, \& {Timmes}}]{2015ApJS..220...15P}
{Paxton}, B., {Marchant}, P., {Schwab}, J., {et~al.} 2015, \apjs, 220, 15

\bibitem[{{Pecaut} \& {Mamajek}(2013)}]{2013ApJS..208....9P}
{Pecaut}, M.~J. \& {Mamajek}, E.~E. 2013, \apjs, 208, 9

\bibitem[{{Peter} {et~al.}(2012){Peter}, {Feldt}, {Henning}, \&
  {Hormuth}}]{2012A&A...538A..74P}
{Peter}, D., {Feldt}, M., {Henning}, T., \& {Hormuth}, F. 2012, \aap, 538, A74

\bibitem[{{Ram{\'{\i}}rez-Tannus} {et~al.}(2017){Ram{\'{\i}}rez-Tannus},
  {Kaper}, {de Koter}, {Tramper}, {Bik}, {Ellerbroek}, {Ochsendorf},
  {Ram{\'{\i}}rez-Agudelo}, \& {Sana}}]{paper1}
{Ram{\'{\i}}rez-Tannus}, M.~C., {Kaper}, L., {de Koter}, A., {et~al.} 2017,
  \aap, 604, A78

\bibitem[{{Ram{\'\i}rez-Tannus} {et~al.}(2020){Ram{\'\i}rez-Tannus}, {Poorta},
  {Bik}, {Kaper}, {de Koter}, {De Ridder}, {Beuther}, {Brandner}, {Davies},
  {Gennaro}, {Guo}, {Henning}, {Linz}, {Naylor}, {Pasquali},
  {Ram{\'\i}rez-Agudelo}, \& {Sana}}]{2020AA...633A.155R}
{Ram{\'\i}rez-Tannus}, M.~C., {Poorta}, J., {Bik}, A., {et~al.} 2020, \aap,
  633, A155

\bibitem[{{Rose} {et~al.}(2019){Rose}, {Naoz}, \&
  {Geller}}]{2019MNRAS.488.2480R}
{Rose}, S.~C., {Naoz}, S., \& {Geller}, A.~M. 2019, \mnras, 488, 2480

\bibitem[{{Sana} {et~al.}(2013){Sana}, {de Koter}, {de Mink}, {Dunstall},
  {Evans}, {H{\'e}nault-Brunet}, {Ma{\'{\i}}z Apell{\'a}niz},
  {Ram{\'{\i}}rez-Agudelo}, {Taylor}, {Walborn}, {Clark}, {Crowther},
  {Herrero}, {Gieles}, {Langer}, {Lennon}, \& {Vink}}]{2013A&A...550A.107S}
{Sana}, H., {de Koter}, A., {de Mink}, S.~E., {et~al.} 2013, \aap, 550, A107

\bibitem[{{Sana} {et~al.}(2012){Sana}, {de Mink}, {de Koter}, {Langer},
  {Evans}, {Gieles}, {Gosset}, {Izzard}, {Le Bouquin}, \&
  {Schneider}}]{2012Sci...337..444S}
{Sana}, H., {de Mink}, S.~E., {de Koter}, A., {et~al.} 2012, Science, 337, 444

\bibitem[{{Sana} \& {Evans}(2011)}]{2011IAUS..272..474S}
{Sana}, H. \& {Evans}, C.~J. 2011, in IAU Symposium, Vol. 272, Active OB Stars:
  Structure, Evolution, Mass Loss, and Critical Limits, ed. C.~{Neiner},
  G.~{Wade}, G.~{Meynet}, \& G.~{Peters}, 474--485

\bibitem[{{Sana} {et~al.}(2008){Sana}, {Gosset}, {Naz{\'e}}, {Rauw}, \&
  {Linder}}]{2008MNRAS.386..447S}
{Sana}, H., {Gosset}, E., {Naz{\'e}}, Y., {Rauw}, G., \& {Linder}, N. 2008,
  \mnras, 386, 447

\bibitem[{{Sana} {et~al.}(2014){Sana}, {Le Bouquin}, {Lacour}, {Berger},
  {Duvert}, {Gauchet}, {Norris}, {Olofsson}, {Pickel}, {Zins}, {Absil}, {de
  Koter}, {Kratter}, {Schnurr}, \& {Zinnecker}}]{2014ApJS..215...15S}
{Sana}, H., {Le Bouquin}, J.-B., {Lacour}, S., {et~al.} 2014, \apjs, 215, 15

\bibitem[{{Sana} {et~al.}(2017){Sana}, {Ram{\'{\i}}rez-Tannus}, {de Koter},
  {Kaper}, {Tramper}, \& {Bik}}]{2017A&A...599L...9S}
{Sana}, H., {Ram{\'{\i}}rez-Tannus}, M.~C., {de Koter}, A., {et~al.} 2017,
  \aap, 599, L9

\bibitem[{{Schneider} {et~al.}(2015){Schneider}, {Izzard}, {Langer}, \& {de
  Mink}}]{2015ApJ...805...20S}
{Schneider}, F.~R.~N., {Izzard}, R.~G., {Langer}, N., \& {de Mink}, S.~E. 2015,
  \apj, 805, 20

\bibitem[{{Sharples} {et~al.}(2013){Sharples}, {Bender}, {Agudo Berbel},
  {Bezawada}, {Castillo}, {Cirasuolo}, {Davidson}, {Davies}, {Dubbeldam},
  {Fairley}, {Finger}, {F{\"o}rster Schreiber}, {Gonte}, {Hess}, {Jung},
  {Lewis}, {Lizon}, {Muschielok}, {Pasquini}, {Pirard}, {Popovic}, {Ramsay},
  {Rees}, {Richter}, {Riquelme}, {Rodrigues}, {Saviane}, {Schlichter},
  {Schmidtobreick}, {Segovia}, {Smette}, {Szeifert}, {van Kesteren}, {Wegner},
  \& {Wiezorrek}}]{2013Msngr.151...21S}
{Sharples}, R., {Bender}, R., {Agudo Berbel}, A., {et~al.} 2013, The Messenger,
  151, 21

\bibitem[{{Sota} {et~al.}(2014){Sota}, {Ma{\'{\i}}z Apell{\'a}niz}, {Morrell},
  {Barb{\'a}}, {Walborn}, {Gamen}, {Arias}, \& {Alfaro}}]{2014ApJS..211...10S}
{Sota}, A., {Ma{\'{\i}}z Apell{\'a}niz}, J., {Morrell}, N.~I., {et~al.} 2014,
  \apjs, 211, 10

\bibitem[{{Sung} {et~al.}(2017){Sung}, {Bessell}, {Chun}, {Yi}, {Naz{\'e}},
  {Lim}, {Karimov}, {Rauw}, {Park}, \& {Hur}}]{2017ApJS..230....3S}
{Sung}, H., {Bessell}, M.~S., {Chun}, M.-Y., {et~al.} 2017, \apjs, 230, 3

\bibitem[{{The Astropy Collaboration} {et~al.}(2018){The Astropy
  Collaboration}, {Price-Whelan}, {Sip{\H o}cz}, {G{\"u}nther}, {Lim},
  {Crawford}, {Conseil}, {Shupe}, {Craig}, {Dencheva}, {Ginsburg},
  {VanderPlas}, {Bradley}, {P{\'e}rez-Su{\'a}rez}, {de Val-Borro}, {Aldcroft},
  {Cruz}, {Robitaille}, {Tollerud}, {Ardelean}, {Babej}, {Bachetti}, {Bakanov},
  {Bamford}, {Barentsen}, {Barmby}, {Baumbach}, {Berry}, {Biscani}, {Boquien},
  {Bostroem}, {Bouma}, {Brammer}, {Bray}, {Breytenbach}, {Buddelmeijer},
  {Burke}, {Calderone}, {Cano Rodr{\'{\i}}guez}, {Cara}, {Cardoso},
  {Cheedella}, {Copin}, {Crichton}, {D{\'A}vella}, {Deil}, {Depagne},
  {Dietrich}, {Donath}, {Droettboom}, {Earl}, {Erben}, {Fabbro}, {Ferreira},
  {Finethy}, {Fox}, {Garrison}, {Gibbons}, {Goldstein}, {Gommers}, {Greco},
  {Greenfield}, {Groener}, {Grollier}, {Hagen}, {Hirst}, {Homeier}, {Horton},
  {Hosseinzadeh}, {Hu}, {Hunkeler}, {Ivezi{\'c}}, {Jain}, {Jenness}, {Kanarek},
  {Kendrew}, {Kern}, {Kerzendorf}, {Khvalko}, {King}, {Kirkby}, {Kulkarni},
  {Kumar}, {Lee}, {Lenz}, {Littlefair}, {Ma}, {Macleod}, {Mastropietro},
  {McCully}, {Montagnac}, {Morris}, {Mueller}, {Mumford}, {Muna}, {Murphy},
  {Nelson}, {Nguyen}, {Ninan}, {N{\"o}the}, {Ogaz}, {Oh}, {Parejko}, {Parley},
  {Pascual}, {Patil}, {Patil}, {Plunkett}, {Prochaska}, {Rastogi}, {Reddy
  Janga}, {Sabater}, {Sakurikar}, {Seifert}, {Sherbert}, {Sherwood-Taylor},
  {Shih}, {Sick}, {Silbiger}, {Singanamalla}, {Singer}, {Sladen}, {Sooley},
  {Sornarajah}, {Streicher}, {Teuben}, {Thomas}, {Tremblay}, {Turner},
  {Terr{\'o}n}, {van Kerkwijk}, {de la Vega}, {Watkins}, {Weaver}, {Whitmore},
  {Woillez}, \& {Zabalza}}]{2018arXiv180102634T}
{The Astropy Collaboration}, {Price-Whelan}, A.~M., {Sip{\H o}cz}, B.~M.,
  {et~al.} 2018, ArXiv e-prints

\bibitem[{{Wellstein} \& {Langer}(1999)}]{1999A&A...350..148W}
{Wellstein}, S. \& {Langer}, N. 1999, \aap, 350, 148

\bibitem[{{Wenger} {et~al.}(2000){Wenger}, {Ochsenbein}, {Egret}, {Dubois},
  {Bonnarel}, {Borde}, {Genova}, {Jasniewicz}, {Lalo{\"e}}, {Lesteven}, \&
  {Monier}}]{2000A&AS..143....9W}
{Wenger}, M., {Ochsenbein}, F., {Egret}, D., {et~al.} 2000, \aaps, 143, 9

\bibitem[{{Whelan} \& {Iben}(1973)}]{1973ApJ...186.1007W}
{Whelan}, J. \& {Iben}, Icko, J. 1973, \apj, 186, 1007

\bibitem[{{Woosley} {et~al.}(1993){Woosley}, {Langer}, \&
  {Weaver}}]{1993ApJ...411..823W}
{Woosley}, S.~E., {Langer}, N., \& {Weaver}, T.~A. 1993, \apj, 411, 823

\bibitem[{{Wu} {et~al.}(2016){Wu}, {Bik}, {Bestenlehner}, {Henning},
  {Pasquali}, {Brandner}, \& {Stolte}}]{2016A&A...589A..16W}
{Wu}, S.-W., {Bik}, A., {Bestenlehner}, J.~M., {et~al.} 2016, \aap, 589, A16

\bibitem[{{Wu} {et~al.}(2014){Wu}, {Bik}, {Henning}, {Pasquali}, {Brandner}, \&
  {Stolte}}]{2014A&A...568L..13W}
{Wu}, S.-W., {Bik}, A., {Henning}, T., {et~al.} 2014, \aap, 568, L13

\bibitem[{{Yoon} {et~al.}(2010){Yoon}, {Woosley}, \&
  {Langer}}]{2010ApJ...725..940Y}
{Yoon}, S.~C., {Woosley}, S.~E., \& {Langer}, N. 2010, \apj, 725, 940

\bibitem[{{Zeidler} {et~al.}(2018){Zeidler}, {Sabbi}, {Nota}, {Pasquali},
  {Grebel}, {McLeod}, {Kamann}, {Tosi}, {Cignoni}, \&
  {Ramsay}}]{2018AJ....156..211Z}
{Zeidler}, P., {Sabbi}, E., {Nota}, A., {et~al.} 2018, \aj, 156, 211

\end{thebibliography}


\begin{appendix}

\section{Accuracy of the age determination}
\label{P5:sec:AccuracyAge}

The effective temperature and bolometric correction of the luminosity class V stars are taken from \citet{2013ApJS..208....9P}; for the luminosity class III stars, we used the calibrations from \citet{1999A&AS..140..261A} for spectral types F0-F9, and for G4-M5 stars we used the calibration of \citet{2000asqu.book.....C}. In order to calculate the luminosity of the observed targets, we used the absolute $K$-band magnitude assuming the \citet{2005ApJ...619..931I} extinction law to derive the $K$-band extinction $A_K$. We then obtained the absolute magnitudes by scaling the apparent magnitudes to the distance of the clusters \citep[$1336_{-76}^{+68}$\,pc for M8 and $1770_{-120}^{+140}$\,pc for \ngc6357;][]{2019ApJ...870...32K}. The luminosities were obtained by using the bolometric correction and the absolute magnitudes. Finally, we determined the age of the regions by comparing the position of the observed stars to MESA isochrones obtained from the MIST project \citep{2011ApJS..192....3P, 2013ApJS..208....4P, 2015ApJS..220...15P, 2016ApJS..222....8D, 2016ApJ...823..102C}. The isochrones and tracks used have a solar metallicity and an initial rotational velocity of $v_{\rm ini}=0.4\,v_{\rm crit}$, the age obtained using tracks with $v_{\rm ini}=0$ are within the reported error bars.

Here we discuss some of the issues that might affect the age determination and compare our age estimates with those published in the literature. 
{\em i)} Binarity could affect the luminosity determination by making stars look brighter than they actually are. Given the spacing of the isochrones at the young age of M8 and \ngc6357, this does not have a significant effect on our age determination. {\em ii)} The use of different evolutionary tracks could lead to different age estimates; for example, \citet{2013A&A...560A..16M} show that the main-sequence age for different models could vary by $\sim0.7~{\rm Myr}$ for a 20~\msun\ isochrone. In our case, this uncertainty is likely smaller than that caused by the uncertainty on the spectral type. {\em iii)} The age determination may depend on the adopted extinction law. By using the \citet{2009ApJ...696.1407N} extinction law, we obtain lower $A_K$ values and, therefore, slightly fainter objects. Nevertheless, given that the extinction in these three regions is relatively modest ($A_{K,{\rm max}} \sim 2.5$), the difference in luminosity produced by the various extinction laws does not significantly affect the age determination. {\em iv)} The classification of the earliest spectral type (O3) is normally degenerate \citep[see e.g.,][]{2014A&A...568L..13W,2016A&A...589A..16W}. In the case of \ngc6357,\ the most massive star is of spectral type O3.5. The fact that its spectral type is uncertain could affect the lower age limit. To avoid this, we determined the age based on the whole population, including the low-mass PMS stars, instead of focusing only on the massive stars. 

\citet{2007MNRAS.374.1253A} estimated the age of M8 to be  $<3$~Myr based on the low-to-intermediate mass PMS population and \citet{2013ApJS..209...26F} estimated the age of the pre-main sequence stars in M8 between $\sim 0.8 - 2.5 ~{\rm Myr}$, which is in agreement with our estimate -- including the massive stars -- of $\minagem-\maxagem~{\rm Myr}$. \citet{2015A&A...573A..95M} determined the age of \ngc6357\ at $1-3~{\rm Myr}$ based on the visual and NIR color-magnitude and color-color diagrams. By comparing the whole population of stars with MIST isochrones, we estimate an age of \minagengc\ to \maxagengc~Myr for this cluster.

We find that the age obtained for the studied regions by looking only at the massive stars is consistent with the age obtained only from the low-mass PMS stars. If the late-K and M-type stars near the Hayashi track were cluster members, we would obtain a younger age ($\sim0.1-1~{\rm Myr}$ younger) for the PMS stars compared to that of the massive stars only.

\section{Lines used to measure the radial velocity}
\label{P5:sec:linesused}

\begin{table}[ht!]
\caption{Radial velocities and lines used to measure the radial velocity for each star in M8.}             
\label{P5:tab:SpecLines_M8}      
\centering                          
\renewcommand{\arraystretch}{1.4}
\setlength{\tabcolsep}{3pt}
\begin{tabular}{cccccccc}        
\hline\hline                 
Name & RV ($\rm km$\,$\rm s^{-1}$) & Br-12 & Br-11 & HeI & Br-10 & HeI & Br$\gamma$ \\
\hline                 
14	 & 	$-$14.1$\pm$7.5	 & 	x 	 & 	x 	 & 	x 	 & 	x 	 & 	x 	 & 	x 	\\
16	 & 	$-$17.4$\pm$8.1	 & 	x 	 & 	x 	 & 		 & 	x 	 & 		 & 	x 	\\
24	 & 	$-$57.1$\pm$7.4	 & 		 & 	x 	 & 	x 	 & 	x 	 & 	x 	 & 	x 	\\
33	 & 	1.3$\pm$16.0	 & 	x 	 & 	x 	 & 		 & 	x 	 & 		 & 	x 	\\
36	 & 	$-$5.1$\pm$7.6	 & 		 & 		 & 	x 	 & 	x 	 & 	x 	 & 	x 	\\
51	 & 	$-$10.2$\pm$9.0	 & 	x 	 & 	x 	 & 	x 	 & 	x 	 & 	x 	 & 	x 	\\
60	 & 	78.4$\pm$12.4	 & 	x 	 & 	x 	 & 	x 	 & 		 & 	x 	 & 	x 	\\
73	 & 	56.6$\pm$12.4	 & 	x 	 & 	x 	 & 		 & 		 & 		 & 	x 	\\
79	 & 	$-$38.1$\pm$10.0	 & 	x 	 & 	x 	 & 	x 	 & 		 & 	x 	 & 		\\
F44	 & 	28.5$\pm$7.2	 & 		 & 		 & 	x 	 & 		 & 	x 	 & 	x 	\\
O10	 & 	11.6$\pm$13.0	 & 	x 	 & 	x 	 & 	x 	 & 	x 	 & 	x 	 & 	x 	\\
O11	 & 	$-$52.9$\pm$35.5	 & 		 & 	x 	 & 		 & 		 & 		 & 	x 	\\
O5	 & 	$-$2.9$\pm$9.0	 & 		 & 		 & 	x 	 & 	x 	 & 	x 	 & 	x 	\\
O6	 & 	29.9$\pm$15.4	 & 	x 	 & 	x 	 & 		 & 	x 	 & 		 & 	x 	\\
O7	 & 	$-$24.5$\pm$13.1	 & 	x 	 & 	x 	 & 	x 	 & 	x 	 & 	x 	 & 	x 	\\
O8	 & 	17.1$\pm$16.6	 & 	x 	 & 	x 	 & 	x 	 & 		 & 	x 	 & 		\\
\hline                                   
\end{tabular}
\end{table}
\begin{table}[ht]
\caption{Radial velocities and lines used to measure the radial velocity for each star in NGC6357.}             
\label{P5:tab:SpecLines_NGC6357}      
\centering                          
\renewcommand{\arraystretch}{1.4}
\setlength{\tabcolsep}{3pt}
\begin{tabular}{cccccccc}        
\hline\hline                 
Name & RV ($\rm km$\,$\rm s^{-1}$) & Br-12 & Br-11 & HeI & Br-10 & HeI & Br$\gamma$ \\
\hline                 
107	 & 	9.5$\pm$14.7	 & 		 & 	x 	 & 		 & 	x 	 & 		 & 		\\
112	 & 	74.8$\pm$9.0	 & 		 & 	x 	 & 	x 	 & 	x 	 & 	x 	 & 	x 	\\
118	 & 	16.5$\pm$13.1	 & 		 & 	x 	 & 	x 	 & 		 & 	x 	 & 	x 	\\
16	 & 	9.8$\pm$8.1	 & 		 & 	x 	 & 	x 	 & 	x 	 & 	x 	 & 	x 	\\
22	 & 	62.1$\pm$16.2	 & 	x 	 & 	x 	 & 	x 	 & 	x 	 & 	x 	 & 	x 	\\
73	 & 	40.4$\pm$11.5	 & 		 & 		 & 	x 	 & 	x 	 & 	x 	 & 	x 	\\
88	 & 	$-$70.1$\pm$12.4	 & 		 & 	x 	 & 	x 	 & 	x 	 & 	x 	 & 	x 	\\
B0	 & 	10.7$\pm$8.3	 & 		 & 	x 	 & 	x 	 & 	x 	 & 	x 	 & 	x 	\\
B10	 & 	32.8$\pm$2.9	 & 		 & 		 & 	x 	 & 	x 	 & 	x 	 & 	x 	\\
B11	 & 	53.9$\pm$7.1	 & 		 & 	x 	 & 	x 	 & 		 & 	x 	 & 	x 	\\
B13	 & 	76.1$\pm$14.4	 & 		 & 	x 	 & 	x 	 & 		 & 	x 	 & 	x 	\\
B15	 & 	38.1$\pm$4.4	 & 		 & 		 & 	x 	 & 		 & 	x 	 & 	x 	\\
B1	 & 	$-$73.6$\pm$12.1	 & 		 & 	x 	 & 	x 	 & 	x 	 & 	x 	 & 	x 	\\
B2	 & 	$-$7.0$\pm$4.0	 & 		 & 		 & 	x 	 & 		 & 	x 	 & 	x 	\\
B4	 & 	$-$7.1$\pm$13.2	 & 		 & 	x 	 & 	x 	 & 		 & 	x 	 & 	x 	\\
B6	 & 	43.3$\pm$13.3	 & 	x 	 & 	x 	 & 		 & 	x 	 & 		 & 	x 	\\
B7	 & 	12.7$\pm$4.1	 & 		 & 		 & 	x 	 & 		 & 	x 	 & 	x 	\\
B8	 & 	$-$41.3$\pm$10.8	 & 		 & 	x 	 & 	x 	 & 	x 	 & 	x 	 & 	x 	\\
B9	 & 	0.5$\pm$2.4	 & 		 & 		 & 	x 	 & 		 & 	x 	 & 		\\
O0	 & 	4.0$\pm$11.0	 & 		 & 		 & 	x 	 & 		 & 	x 	 & 	x 	\\
O1	 & 	36.7$\pm$5.3	 & 		 & 		 & 	x 	 & 		 & 	x 	 & 	x 	\\
O2	 & 	35.6$\pm$10.1	 & 	x 	 & 		 & 	x 	 & 	x 	 & 	x 	 & 	x 	\\
\hline                                   
\end{tabular}
\end{table}


\section{Distribution of radial velocities of the OB stars in M8 and \ngc6357}
\label{P5:sec:RVhistograms}

\begin{figure}[ht]
\setlength{\tabcolsep}{-3pt}
     \centering
        \subfigure[M8]{%
           \includegraphics[width=0.5\hsize]{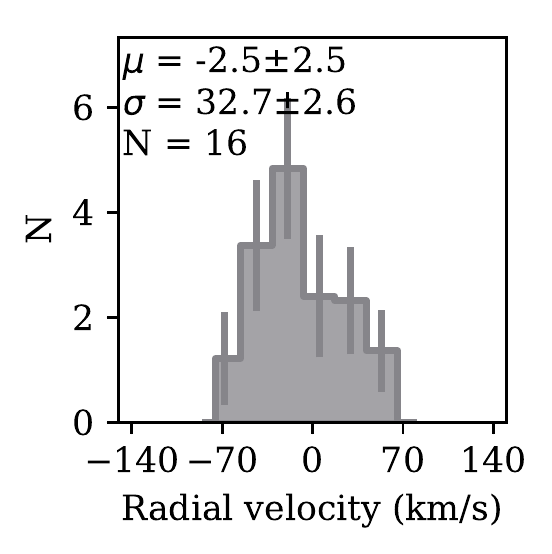}
        }\hspace*{-0.9em}%
        \subfigure[\ngc6357]{%
           \includegraphics[width=0.5\hsize]{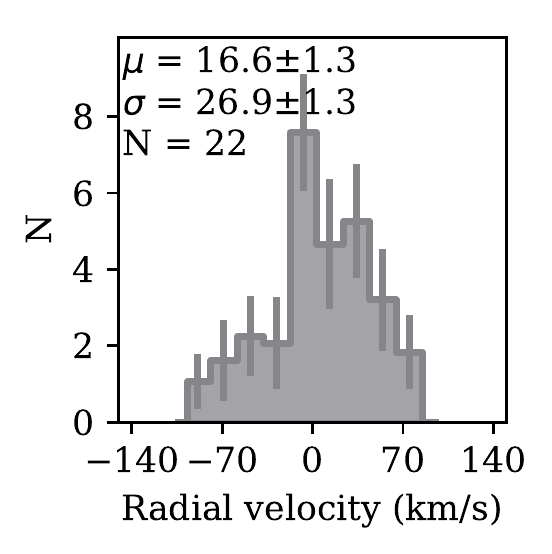}
        }\\%
    \caption[]{Distribution of radial velocities for the massive stars in M8 and \ngc6357. 
    }
\label{P5:fig:RV_histograms}
\end{figure}


\section{Comparison to previous work}
\label{P5:sec:PrevWork}

We compare the \srv\ distribution obtained with the distribution presented in \citet{2017A&A...599L...9S}. Therefore, we sampled clusters of 12 stars with \pcutoff$ = 1.4$~days and \fbin$ = 0.7$ and the radial velocity measurement errors of M17. This results in a \srv$ = 40.9^{+18.9}_{-16.1}$~\kms\ for the method described above. \citet{2017A&A...599L...9S} found a lower value of \srv$ = 35.0^{+21.9}_{-14.9}$~\kms. The confidence intervals of the two values overlap due to the width of the distribution.

We assessed the correctness of our method by generating a parent population using numerically calculated two body orbits. The binary star systems were generated in a similar way as described above. Each system is described by a primary mass, mass ratio, period, and eccentricity using identical distributions as before. The orbits of the binaries were simulated in a two dimensional plane. Only a single orbit starting at periastron was simulated numerically using a fourth order Runge-Kutta method \citep{dormand1980family}. The systems were tested on their stability by integrating for a large number of orbits. The velocity was obtained by selecting a random time step from the orbit. The two dimensional $x$ and $y$ velocities were converted to a radial velocity using a randomly generated inclination and longitude of periastron. Finally, the effect of cluster dynamics was added to the radial velocity. 

The radial velocities calculated from the numerical two body orbits were combined with the radial velocities of single stars to generate a new parent population of $10^5$ stars. This parent population can be sampled as described above to generate a \srv\ distribution. 
Using the same cluster properties, we produced a \srv\ distribution for M17 with \fbin$=0.7$ and \pcutoff$=1.4$~days. This resulted in \srv$ = 40.5^{+18.6}_{-16.0}$~\kms, which agrees closely with this work. No significant difference with this work was found for any binary fraction or cutoff period.

\end{appendix}

\end{document}